\documentclass[preprint,showpacs,preprintnumbers,superscriptaddress,amsmath,amssymb]{revtex4-1}

\usepackage{graphicx}
\usepackage{colortbl}
\usepackage{mathrsfs}
\usepackage{bm} 
\usepackage{ulem}

\usepackage{amsmath}

\renewcommand{\Re}{\operatorname{Re}}
\renewcommand{\Im}{\operatorname{Im}}

\makeatletter
\def\btt#1{\texttt{\@backslashchar#1}}%
\DeclareRobustCommand\bblash{\btt{\@backslashchar}}%
\makeatother

\begin{document}


\title{Energetic perspective on emergent inductance exhibited by magnetic textures in the pinned regime}

\author{Soju Furuta}
\affiliation{Department of Physics, Tokyo Institute of Technology, Tokyo 152-8551, Japan}

\author{Samuel Harrison Moody}
\affiliation{Durham University, Centre for Materials Physics, Durham, DH1 3LE, United Kingdom}

\author{Kyohei Kado}
{\affiliation{Department of Applied Physics, The University of Tokyo, Tokyo 113-8656, Japan}

\author{Wataru Koshibae}
\affiliation{RIKEN Center for Emergent Matter Science (CEMS), Wako 351-0198, Japan}

\author{Fumitaka Kagawa}
\email{kagawa@phys.titech.ac.jp}
\affiliation{Department of Physics, Tokyo Institute of Technology, Tokyo 152-8551, Japan}
\affiliation{RIKEN Center for Emergent Matter Science (CEMS), Wako 351-0198, Japan}

\date{\today}
\begin{abstract}
Spatially varying magnetic textures can exhibit electric-current-induced dynamics as a result of the spin-transfer torque effect. When such a magnetic system is electrically driven, an electric field is generated, which is called the emergent electric field. In particular, when magnetic-texture dynamics are induced under the application of an AC electric current, the emergent electric field also appears in an AC manner, notably, with an out-of-phase time profile, thus exhibiting inductor behaviour, often called an emergent inductor. Here we show that the emergent inductance exhibited by magnetic textures in the pinned regime can be explained in terms of the current-induced energy stored in the magnetic system. We numerically find that the inductance values defined from the emergent electric field and the current-induced magnetization-distortion energy, respectively, are in quantitative agreement in the so-called adiabatic limit. Our findings indicate that emergent inductors retain the basic concept of conventional inductors; that is, the energy is stored under the application of electric current.
\end{abstract}

\maketitle

Strong coupling between conduction-electron spin and underlying spin texture forms the basis of rich phenomena, such as electric-current-induced dynamics of the spin system via the spin-transfer-torque (STT) effect \cite{STT1, STT2, YamanouchiNature} and the spin-dynamics-induced electromotive force (spin motive force or emergent-electric field (EEF)) \cite{Volovik, BarnesPRL2007, YangPRL, JiadongPRL, Schulz_Nat.Phys.}. Although obtaining a general expression for the EEF is difficult, a concise form is available for a specific situation; that is, the magnetic texture is slowly varying in space, and the conduction-electron spins are always parallel to the local magnetic moments of the magnetic texture. In this limit, often referred to as the adiabatic limit, electron transport under the influence of the magnetic texture is described by introducing the effective U(1) gauge field, which results in the emergent magnetic field and EEF \cite{Volovik}. The EEF, which is the focus of this study, can be described in the following equation \cite{Volovik, BarnesPRL2007}:
\begin{equation}
\label{e-field}
e_i(\bm{r}, t) = \frac{\hbar}{2|e|} \bm{m}(\bm{r}, t) \cdot [{\partial}_i \bm{m}(\bm{r}, t) \times {\partial}_t \bm{m}(\bm{r}, t)],
\end{equation}
where $e (>0)$ is the elementary charge, $\bm{m}(\bm{r}, t)$ is the unit vector of the local magnetic moment at position $\bm{r}$ and time $t$, and ${\partial}_i$ ($i = x, y, z$) and ${\partial}_t$ denote the spatial and time derivatives, respectively. When the conduction-electron spins are not fully polarized, the so-called spin-polarization factor $P$ is further considered for the resulting electric field \cite{Schulz_Nat.Phys., BarnesPRL2007}. As explicitly expressed in this equation, the EEF can appear only when the magnetic texture is time evolving.

Recently, the interplay between the STT and EEF has attracted much attention as a source of a new class of inductor, often called an emergent inductor \cite{NagaosaJJAP, YokouchiNature, KitaoriPNAS, IedaPRB, Kurebayashi, KitaoriPRB}. To understand the emergent inductor under the application of an AC electric current, it is still instructive to consider the dynamics of a magnetic system under a DC electric current.
In the following, we focus on the so-called pinned regime \cite{Nattermann, ChauvePRB, Kleemann}, in which a magnetic system does not exhibit a steady flow under a DC electric current \cite{IntrinsicTatara, ThiavilleEPL, ExtrinsicTatara, IntrinsicOno, ExtrinsicNatPhys,Iwasaki_Nat.Commun., TataraReview}.
When a DC current (let $j$ be the current density) is applied, a magnetic texture starts to deform as a result of the STT effect, but its dynamics are only transient and eventually stop at $t \rightarrow \infty$ by definition of the pinned regime; thus, ${\partial}_t \bm{m} = 0$ at the final state, and hence, no EEF appears: We will illustrate the case of a helical magnetic texture in the Results section. In the linear-response regime, the change in the local magnetization direction at the final state, $\delta$$\bm{m}$, is elastic and proportional to $j$; i.e., $\delta \bm{m} \propto j$.

The situation under an AC electric current, $j(t) = j_0 \sin \omega t$, can be considered in a similar way. As long as the linear and low-frequency response regimes are considered, such that $\delta \bm{m}(t)$ is proportional to the instantaneous value of $j(t)$. In this limit, the magnetization has an in-phase response to the applied AC current: $\delta \bm{m}(t) \propto j(t) = j_0 \sin \omega t$, Note that as a natural consequence of the application of an AC current, the current-induced dynamics persist, and thus ${\partial}_t \delta \bm{m}$ remains finite even in the pinned regime; ${\partial}_t \delta \bm{m} \propto j_0 \omega \cos \omega t = {\rm d}j(t)/{\rm d}t $, and this is an out-of-phase response to the applied AC current. Thus, an out-of-phase linear-response EEF can appear because $\bm{m} \cdot ({\partial}_i \bm{m} \times {\partial}_t \delta \bm{m})$ is finite, and it is expressed as:
\begin{equation}
\label{dj/dt}
e_i(t) = \tilde{L}\frac{{\rm d}j(t)}{{\rm d}t},
\end{equation}
where $\tilde{L}$ is a normalized inductance (the unit is henry metre, H m, which we may term ``inductivity'', in analogy to the terminology of resistivity) defined in the linear-response and low-frequency regimes. By multiplying the sample length $\ell$ with both sides of Eq.~(\ref{dj/dt}) and inserting $I = Aj$, where $I$ and $A$ respectively represent the applied current and the sample cross-section area, one can obtain a standard equation describing the self-induction phenomenon:
\begin{equation}
\label{dI/dt}
V = L\frac{{\rm d}I}{{\rm d}t},
\end{equation}
where $L\equiv \tilde{L}(\ell /A)$ and $V$ represent the self-induction coefficient and the inductive counter-electromotive force, respectively. This voltage-current relation in the inductor defines $L$. In general, $L$ is frequency dependent and may be represented by a complex number, $L^{*}(\omega)$, the imaginary part of which describes the phase delay of the voltage response from ${\rm d}I/{\rm d}t$.

Nevertheless, as long as one considers the low-frequency regime such that $\Re L^{*}(\omega) \gg \Im L^{*}(\omega)$ is satisfied, the inductance can be taken as a constant real number. In this case, the electric work required to supply a current to the inductor (assume $I=0$ for $t \leq 0$) is calculated from Eq.~(\ref{dI/dt}) as follows:
\begin{eqnarray}
\label{integral}
\int_0^{t}{\rm d}t'I(t')V(t') &=& \int_0^{t}{\rm d}t'\frac{{\rm d}}{{\rm d}t'}\left( \frac{1}{2}LI(t')^2 \right) \nonumber \\
&=& \frac{1}{2}LI(t)^2.
\end{eqnarray}
This electric work, $\frac{1}{2}LI(t)^2$, should be positive, and as is clear from the derivation, it is nondissipative in nature; thus, given energy conservation, the corresponding energy can be viewed as stored in the inductor in the circuit. Alternatively, the electric work done by the external power supply can also be viewed as stored in the energy of the whole system; thus, the energy increase of the whole system, $\Delta E_{\rm system}(I(t))$, satisfies $\Delta E_{\rm system}(I(t)) = \frac{1}{2}LI(t)^2$.

In a classical inductor made of a coil, it can be shown analytically that $\Delta E_{\rm system}(t)$ is equal to the magnetic-field energy, $\frac{1}{2} \int {\rm d}V \bm{H}(t) \cdot \bm{B}(t)$. Thus, at least for the case of the classical inductors, the value of $L$ in the low-frequency regime can be defined in two ways: One is from the electric response due to the electromagnetic induction (EMI), and $L_{\rm EMI}$ is given by $ V(t) = L_{\rm EMI}\frac{{\rm d}I(t)}{{\rm d}t}$; the other is from the energy increase of the whole system, and $L_{\rm energy}$ is given by $\Delta E_{\rm system}(t) =  \frac{1}{2}L_{\rm energy}I(t)^2$. Although the two definitions of $L$ are based on different perspective, they result in the same value, $L_{\rm EMI} = L_{\rm energy}$.

In the case of emergent inductors, the microscopic mechanism is based on the quantum mechanics, and it is thus quite different from that of classical inductors. Nevertheless, Eq.~(\ref{dI/dt}) remains valid for describing the electric response of emergent inductors, and energy conservation should invariably hold. Therefore, it is expected that the value of $L$ of the emergent inductor in the low-frequency regime can be defined also in terms of $\Delta E_{\rm system}$. 
However, to the best of the authors' knowledge, the emergent inductance has never been discussed from such an energetic perspective, although the energetic implications of Eq.~(\ref{e-field}) has been discussed by Barnes and Maekawa \cite{BarnesPRL2007}.

In this paper, we address this issue numerically using micromagnetic software. We consider a magnetic system in a pinned regime and perform micromagnetic simulations for the current-induced dynamics of a magnetic texture; then, we calculate the time evolutions of the voltage due to EEF, $V_{e}(t)$, and the magnetic-system energy, $\Delta E(t)$, according to Eq.~(\ref{e-field}) and our model spin-Hamiltonian (see below), respectively. We can thus numerically derive $L_{\rm EMI}$ and $L_{\rm energy}$ and, by comparing the two values, test the energetic perspective on the emergent inductor.

By investigating the energetic perspective, we also aim to gain insight into the meaning of negative emergent inductance, an intriguing issue reported in the past experimental studies \cite{YokouchiNature, KitaoriPNAS, KitaoriPRB} and theoretical studies \cite{IedaPRB, Kurebayashi}. The term ``negative inductance'' immediately evokes many questions: Is the negative emergent inductor really stable, even though the negative inductance is known to be unstable (Supplementary Note 1)? Similarly, does the negative inductance mean that the current-supplied state of the material have a lower energy than that of the zero-current state? Or does the emergent inductance no longer have an energetic meaning, even though electrodynamics textbooks state that  the energy definition is fundamental for inductance \cite{Jackson}? It is known that complex impedance at low frequencies can be analyzed by assuming an appropriate equivalent circuit consisting of positive, real-valued circuit elements, $R$ (resistance), $L$ (inductance), and $C$ (capacitance). However, in what cases does negative $L$ need to be introduced beyond this well-established framework? Since these fundamental questions remain unanswered, the physical meaning of the negative inductance remains unclear.

\section{R\lowercase{esults}}
\subsection{Models}
Simulating an emergent inductance using micromagnetic software has the following limitations.
First, in calculating the magnetic-texture dynamics under the electric current and resulting EEF, we should take the spatial derivative of the magnetic texture, ${\partial}_i \bm{m}(\bm{r})$. It follows that, for our approach to be valid, the magnetic texture that we consider should be slowly varying in space. To minimize this problem, in this study, we restrict ourselves to long-period helical magnetic textures.

Second, although the real system is ultimately an electron-spin-coupled system, we consider a Hamiltonian and an equation of motion, both of which describe the magnetic subsystem only. Thus, we can calculate the current-induced energy increase $\Delta E(t)$ only for the local magnetic moments, which is not, strictly speaking, equal to $\Delta E_{\rm system}(t)$ when an energy increase in the electronic subsystem is not negligible.

Third, our calculation of the EEF is based on Eq.~(\ref{e-field}). As long as a slowly varying magnetic texture is considered, one can describe electron transport properties by introducing a gauge field, which has in general SU(2) symmetry \cite{TataraReview}. By taking the adiabatic limit, the SU(2) gauge field reduces to a U(1) gauge field: Eq.~(\ref{e-field}) is thus derived. Conversely, when the system deviates from the adiabatic limit, the use of Eq.~(\ref{e-field}) becomes less justified.

Note that the first issue is, in principle, avoidable by considering a sufficiently slowly varying magnetic texture and increasing numerical efforts. In contrast, the other two issues are more fundamental and thus unavoidable as long as the approach is based on the spin-only Hamiltonian and Eq.~(\ref{e-field}), which is the formalism for the adiabatic limit.

In this study, we consider long-period helical magnetic textures that are stabilized by the Dzyaloshinskii-Moriya (DM) interaction. We consider both a clean system without any disorder and dirty systems including randomly distributed disorder. Our model Hamiltonian is:
\begin{eqnarray}
\label{Hamiltonian}
\begin{split}
\mathscr{H} = &\int \frac{{\rm d}^3r}{a^3} \left[ \frac{J}{2}(\nabla\bm{m})^2 + D\bm{m}\cdot(\nabla\times\bm{m}) \right] \\
&- \sum_{k \in \Lambda} \int_{V_k} {\rm d}^3r \: K_{{\rm imp}} (\bm{m}_k \cdot \bm{n}_{{\rm imp}, k})^2
\end{split}
\end{eqnarray}
where $J$ is the exchange stiffness, $D$ is the DM interaction and $a$ is the lattice constant. When examining a randomness effect, we introduce the last term of Eq.~(\ref{Hamiltonian}): $K_{{\rm imp}} (>0)$ represents the magnetic-easy-axis anisotropy along a randomly chosen direction, $\bm{n}_{{\rm imp}, k}$, at the $k$-th cell (the cell volume $V_k$ is $3^3$ nm$^3$), and $\Lambda$ is a set of random numbers.

When simulating the current-induced dynamics of a given helical magnetic structure, we insert the spin Hamiltonian into the following Landau-Lifshitz-Gilbert (LLG) equation \cite{LLG}:
\begin{eqnarray}
\label{LLG}
\begin{split}
\frac{d\bm{m_r}(t)}{dt}=-\frac{\gamma}{1+\alpha^2} \frac{{\rm d}\mathscr{H}}{{\rm d}\bm{m_r}} &\times \bm{m_r} + \frac{\alpha \gamma}{1+\alpha^2}\left[\bm{m_r}\times \left(\frac{{\rm d}\mathscr{H}}{{\rm d}\bm{m_r}} \times \bm{m_r} \right)\right]\\
+ \frac{1}{1 +\alpha^2}\{ (1&+\beta \alpha)\bm{m_r} \times [\bm{m_r}\times (\bm{u}\cdot \bm{\nabla})\bm{m_r}]\\
&- (\beta - \alpha)[\bm{m_r}\times (\bm{u}\cdot \bm{\nabla})\bm{m_r}] \},
\end{split}
\end{eqnarray}
where $\bm{u}$ represents the spin drift velocity, $\alpha$ is the Gilbert damping constant, $\beta$ is a dimensionless constant that characterizes the nonadiabatic electron spin dynamics, and $\gamma (>0)$ is the gyromagnetic ratio;
$\bm{u}$ is related to the electric current density $\bm{j}$ by $\bm{u} = \frac{P\mu_{\rm B}}{2eM_{\rm s}(1+\beta^2)}\bm{j}$, where $\mu_{\rm B}$ is the Bohr magneton and $M_{\rm s}$ is the saturation magnetization.
When implementing the micromagnetic simulation, we use the open software MuMax3 \cite{Mumax1, Mumax2}.
We choose the following parameter set:
$J/(2a^3) = 1.8\times10^{-11}$ J m$^{-1}$, $D/a^3 = 2.8\times10^{-3}$ J m$^{-2}$, $K_{{\rm imp}} = 1.0\times 10^6$ J m$^{-3}$, $M_{\rm s} =2.45\times10^5$ A m$^{-1}$, $P = 1$, and $\alpha = 0.04$.
In the following simulation, we apply a current density of a sufficiently small magnitude so that the magnetic system is certainly in a pinned regime.

As shown below, we find that with respect to the input AC electric current, $j(t) = j_0\sin \omega t$, the output AC emergent voltage, $V_e(t)$ is $\propto j_0\omega \cos \omega t$, and the time-evolving magnetic-system energy, $\Delta E(t)$ is $\propto (j_0\sin \omega t)^2$. From these observations, $L_{\rm EMI}$ and $L_{\rm energy}$ are derived from the following equations:
\begin{eqnarray}
\label{e-volt}
V_e(t) = \langle e_x(t) \rangle \ell  = L_{\rm EMI} \frac{{\rm d}I(t)}{{\rm d}t} = \left(\tilde{L}_{\rm EMI} \frac{\ell}{A} \right) \frac{{\rm d}I(t)}{{\rm d}t},\\
\label{energy}
\Delta E(t) =\frac{1}{2}L_{\rm energy}I(t)^2 = \frac{1}{2}\left(\tilde{L}_{\rm energy} \frac{\ell}{A} \right) I(t)^2,
\end{eqnarray}
where $\langle \cdots \rangle$ denotes a spatially averaged value.

\subsection{Clean systems}

As one of the simplest systems, we first study the following quasi-one-dimensional system: the system size is $243\times18\times1$ ($\ell =243\times3$ nm and $A=18\times 1\times 3^2$ nm$^2$) under the periodic boundary condition, including no disorder. In such a clean system, the spin system exhibits a pristine helical structure with the $yz$ helical plane (Fig.~1a, which illustrates the case of right-handed chirality). When $\beta = 0$ in Eq.~(\ref{LLG}) and the electric current is below a threshold value (in the present system, $\approx$5.0$\times$10$^{12}$ A m$^{-2}$), the spin system is in the so-called intrinsic pinning regime \cite{ThiavilleEPL, IntrinsicOno, TataraReview}; namely, when a DC current is applied at $t=0$, the helical texture starts to deform along the current direction, and after $\sim$1 ns, static and elastic tilting along the $x$ direction is realized, forming a conical state with a net magnetization (Figs.~1b--d): In addition, the position at which the local magnetic moment exhibits the maximum $m_z$ is slightly displaced and stopped (Fig.~1e). In contrast, when $\beta$ is finite, the helical texture exhibits a steady flow for arbitrary small current density because of the absence of any disorder \cite{IntrinsicTatara, ThiavilleEPL, Iwasaki_Nat.Commun.}, and thus in the clean system, the magnetic-texture dynamics in a pinned regime, which is a focus of this study, is realized only when $\beta = 0$. When an AC electric current is applied, magnetic moment tilting occurs within the pinned regime toward the $+$$x$ and $-$$x$ directions alternatingly with time, yielding an alternating electric field according to Eq.~(\ref{e-field}).

\begin{figure*}
\includegraphics{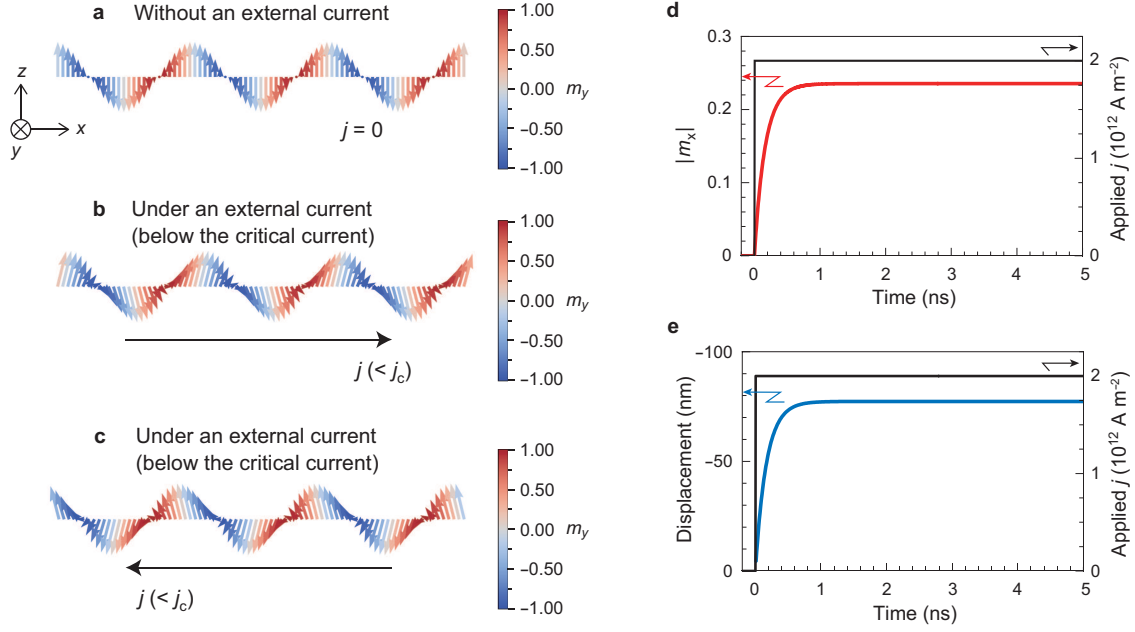}
\caption{\label{} \textbf{Schematics for a helical magnetic texture and its distortion under the application of an electric current.} \textbf{a} Pristine helical magnetic structure with right-handed chirality under zero electric current. \textbf{b, c} Illustration of the current-induced distortion of the helical magnetic structure under rightward (\textbf{b}) and leftward (\textbf{c}) electric currents at steady state. \textbf{d} Time profile of $m_x$ under a DC current application. The tilting direction of helical magnetism shown in \textbf{b} and \textbf{c} is reversed for a helical magnetic structure with left-handed chirality (not shown). \textbf{e} Time profile of the translational displacement, which is defined with respect to the position at which the local moment exhibits the maximum $m_z$. In \textbf{b--e}, a relatively large current density, 2.0$\times10^{12}$ A m$^{-2}$, is used to obtain a large distortion, just for clarity, but it is still lower than the critical current density, $\approx$5.0$\times$10$^{12}$ A m$^{-2}$. In the present case, $m_x$ is uniform in space.}
\end{figure*}

For such a pristine helical magnetic texture, the current-induced dynamics in the pinned regime can be analytically derived within the framework of Eq.~(\ref{LLG}) with $\beta = 0$. Thus, assuming Eq.~(\ref{e-field}) for the EEF, the microscopic expression for $L_{\rm EMI}$ can be derived as reported in previous theoretical studies \cite{NagaosaJJAP, IedaPRB} (see also Supplementary Note 2): The result is
\begin{eqnarray}
\label{L_EMI}
    L_{\mathrm{EMI}} =  \left(  \frac{P\hbar }{2|e|} \right) ^2 \frac{a^3}{J} \frac{\ell}{A} = \tilde{L}_{\mathrm{EMI}} \frac{\ell}{A}.
\end{eqnarray}
Similarly, the energy increase, $\Delta E$, can also be derived (for the derivation, see Supplementary Note 2); then, by assuming the energy conservation ($\Delta E = \frac{1}{2}L_{\rm energy}I^2$), the expression of $L_{\rm energy}$ can be obtained:
\begin{eqnarray}
\label{L_energy}
    L_{\mathrm{energy}} =  \left(  \frac{P\hbar }{2|e|} \right) ^2 \frac{a^3}{J} \frac{\ell}{A} = \tilde{L}_{\mathrm{energy}} \frac{\ell}{A}.
\end{eqnarray}
Thus, we find $\tilde{L}_{\mathrm{EMI}} = \tilde{L}_{\mathrm{energy}}$ analytically for the case of the intrinsically pinned helical magnetic texture ($\beta = 0$). This agreement means that the energy increase of the magnetic system is equal to the work done by the external power supply against the inductive counter-electromotive force due to the EEF, and is consistent with the first law of thermodynamics. By substituting $J/(2a^3) = 1.8\times10^{-11}$ J m$^{-1}$ into Eqs.~(\ref{L_EMI}) or (\ref{L_energy}), we obtain $\tilde{L} =3.006 \times 10^{-21}$ H m. This value can be used to test the validity and accuracy of our numerical approach.

\begin{figure}
\includegraphics{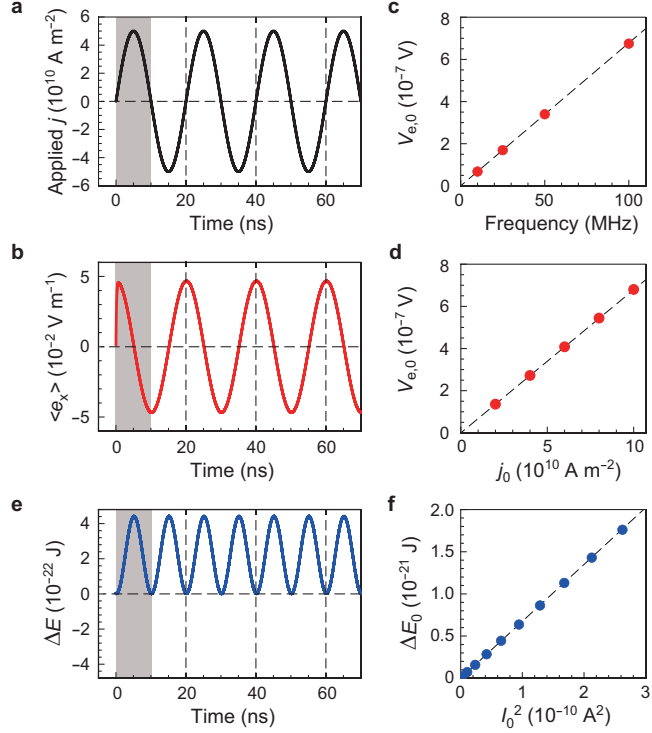}
\caption{\label{} \textbf{Electric and energetic responses emerging from AC-current-induced dynamics of the helical magnetic texture.} \textbf{a, b} Time profiles of the applied AC current (\textbf{a}) and resulting emergent electric field (\textbf{b}). \textbf{c, d} Frequency (\textbf{c}) and current amplitude (\textbf{d}) dependences of the amplitude of the emergent electric field. \textbf{e} Time profile of the magnetic-system energy. \textbf{f} Amplitude of the oscillating magnetic-system energy as a function of current square, $I_0^{2}$. In \textbf{a, b, e}, the data in the first half cycle (gray hatched) are excluded from the analysis to analyze a system that is sufficiently settled for a steady-state cycle under an AC current. These results are obtained in the clean system, and the qualitatively same results are also obtained in the dirty systems.}
\end{figure}

To numerically derive the value of the emergent inductance, we calculate $V_{\rm e}(t)$ and $\Delta E(t)$ for the AC current-induced helical-texture dynamics. When $j(t) = j_0 \sin \omega t$ with $j_0 = 5.0\times 10^{10}$ A m$^{-2}$ and $\omega /2\pi = 50$ MHz is applied (Fig.~2a), the emergent voltage exhibits $V_{\rm e}(t) = V_{\rm e, 0} \cos \omega t$ (Fig.~2b). We further confirm that $V_{\rm e, 0}$ is proportional to both $\omega$ and $j_0$ (Fig.~2c and d, respectively). These observations are consistent with the behaviour expressed by Eq.~(\ref{e-volt}), representing a numerical demonstration of the emergent inductor consisting of a helical magnetic structure. We also find that $\Delta E(t)$ changes according to $\Delta E(t) = \Delta E_{0}\sin^{2} \omega t$ (Fig.~2e), and the amplitude $\Delta E_{\rm 0}$ is proportional to $I_0^2 = (Aj_0)^{2}$ (Fig.~2f), consistent with $\Delta E(t) \propto j(t)^2$, as expressed by Eq.~(\ref{energy}). From these behaviours, we obtain $\tilde{L}_{\rm EMI} = 2.98 \times 10^{-21}$ H m and $\tilde{L}_{\rm energy} = 2.96 \times 10^{-21}$ H m. The relative error $\delta$, defined by $\delta = \vert \tilde{L}_{\rm EMI} - \tilde{L}_{\rm energy} \vert/\tilde{L}_{\rm energy}$, is less than 1 $\%$, leading us to conclude that $\tilde{L}_{\rm EMI} = \tilde{L}_{\rm energy}$ is confirmed within the numerical error. Furthermore, these numerical results have an error of less than 2$\%$ within the theoretical value, $\tilde{L} =3.006 \times 10^{-21}$ H m, supporting the validity of our numerical approach.

\subsection{Dirty systems}

\begin{figure}
\includegraphics{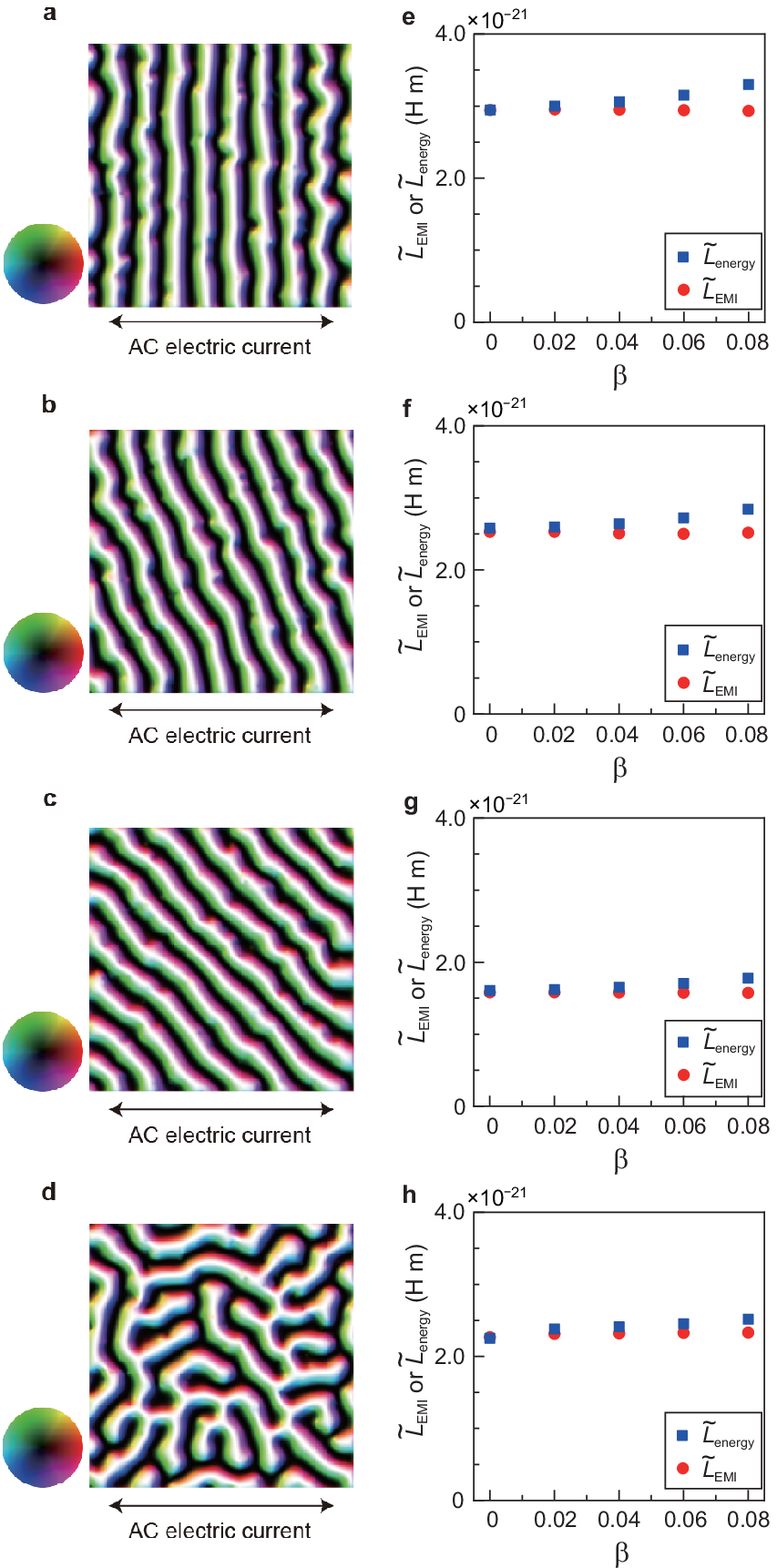}
\caption{\label{} \textbf{Various metastable helical textures (a--d) and corresponding inductivity (e--h).} The current-induced magnetic texture dynamics are calculated under the application of a weak AC current, $5.0 \times 10^{10}$ A m$^{-2}$. This current magnitude is so weak that the current-induced deformation of the magnetic textures is not discernible in the images, but it is numerically detectable. The frequency of the AC current is 50 MHz, which is so low that the response delay of the texture deformation from the time-varying current is negligible. These conditions are satisfied for all $\beta$ values we investigated in this study ($\alpha = 0.04$). To obtain the metastable helical textures at $t=0$, a pristine helical texture with a different oblique angle of the helical $\bm{q}$-vector (for \textbf{a--c}) or a random spin configuration (for \textbf{d}) is prepared as an initial state and then relaxed under zero current.}
\end{figure}

To see the universality of $L_{\rm EMI} = L_{\rm energy}$, it is helpful to numerically examine disordered helical textures in dirty systems. To this end, we prepare a two-dimensional system (the system size is $243 \times 243 \times 1$; i.e., $\ell =243 \times 3$ nm and $A=243\times 1\times 3^2$ nm$^2$), introduce the disorder cells (the density is 3 $\%$), and impose open-boundary conditions. In such a dirty system, the helical textures remain in a pinning regime even for finite $\beta$ (often referred to as an extrinsic pinning regime \cite{ThiavilleEPL, ExtrinsicTatara, ExtrinsicNatPhys}) as long as the applied electric current is below a threshold value (in the present system, $\approx$5.0$\times$10$^{12}$ A m$^{-2}$ for $\beta=0$ and $\approx$1.5--2.0$\times$10$^{12}$ A m$^{-2}$ for finite $\beta$ ($0.02 \leq \beta \leq 0.08$)); hence, in dirty systems, $L_{\rm EMI}$ and $L_{\rm energy}$ exhibited by the magnetic-texture dynamics in a pinned regime can be examined for both $\beta = 0$ and $\beta > 0$. Note that because of the presence of disorder, the spin texture can adopt various metastable states. Here, we show four different examples of metastable helical textures, each of which are shown in Fig.~3a--d: The helical $\bm{q}$-vector of the three systems (Fig.~3a--c) forms approximately 0, 20, and 45 degrees with the AC current direction along the horizontal direction, respectively, whereas the highly disordered helix shown in Fig.~3d has no specific $\bm{q}$-vector.

For the dirty systems, similar to the case of the clean system, we obtain $V_e(t) \propto \frac{{\rm d}(Aj(t))}{{\rm d}t}$ and $\Delta E(t) \propto (Aj(t))^2$; thus, $\tilde{L}_{\rm EMI}$ and $\tilde{L}_{\rm energy}$ are derived separately.
Figure 3e--h summarize the results for the four systems. We find that for all magnetic systems, $\tilde{L}_{\rm EMI} \approx \tilde{L}_{\rm energy}$ invariably holds within 2 $\%$ relative error for $\beta = 0$, whereas such a good agreement is not seen for finite $\beta$. Parenthetically, among the three helical structures shown in Fig.~3a--c, the inductivity is maximized when the helical $\bm{q}$-vector is parallel to the current direction, consistent with the fact that the STT is most effective when the current is along the magnetic modulation direction.

\section{D\lowercase{iscussion}}

Our numerical observations suggest that as long as a slowly varying magnetic texture in a pinned regime is considered, the limitations discussed in the Models section play a minor role at $\beta = 0$. The implications of these observations are that at $\beta = 0$, (i) the EEF can be well described by Eq.~(\ref{e-field}), which is the formalism derived in the adiabatic limit, and (ii) the EEF described by Eq.~(\ref{e-field}) is also consistent with the current-induced magnetic-texture dynamics described by Eq.~(\ref{LLG}) in terms of the energy conservation. Although there is some controversy about the physical meaning of $\beta$ \cite{LLG, BarnesPRL2005, JPSJKohno, DuinePRB, TserkovnyakPRB}, some microscopic approaches indicate that the adiabatic limit corresponds to $\beta = 0$ \cite{LLG, JPSJKohno, TserkovnyakPRB}, and it was discussed \cite{TserkovnyakPRB, DuinePRB2, ShibataPRB, YamaneJMMM} that Eq.~(\ref{e-field}) is valid only for $\beta=0$. Our numerical observations appear to be consistent with this theoretical argument.

In the three dirty systems with a different oblique angle of the helical $\bm{q}$-vector and one highly disordered helical texture in which the $\bm{q}$-vector is ill-defined, we observe a tendency that as $\beta$ increases, the agreement between $\tilde{L}_{\rm EMI}$ and $\tilde{L}_{\rm energy}$ worsens and $\tilde{L}_{\rm energy}$ becomes larger than $\tilde{L}_{\rm EMI}$; i.e., for finite $\beta$, the increase of the magnetic system energy due to current exceeds the work done by the external power supply, and the present framework does not conserve energy [but the relative error is still less than 12 $\%$ at the highest $\beta (=0.08)$ for the magnetic textures considered in this study]. Thus, it appears that in order to satisfy the energy conservation at finite $\beta$, the EEF must be greater than that given by Eq.~(\ref{e-field}). In this context, we note that several theoretical studies have led to an additional correction term, $-\beta \frac{\hbar}{2|e|} \left({\partial}_t \bm{m} \cdot {\partial}_i \bm{m}  \right)$, on the right-hand side of Eq.~(\ref{e-field}), that was derived using a different perspective \cite{TserkovnyakPRB, DuinePRB2, ShibataPRB, YamaneJMMM}. However, it can be shown both analytically and numerically that adding this term further decreases $\tilde{L}_{\rm EMI}$ (Supplementary Note 2 and Fig.~S1); in fact, it was the contribution of this correction term that led to the possibility of negative inductance in the previous theoretical study \cite{IedaPRB}. It remains a challenge for the future to establish a theoretical framework that self-consistently describes the energy and EEF associated with current-induced magnetic-texture dynamics, especially for finite $\beta$.

Given these things, it appears also challenging to quantitatively describe the EEF, for instance, in nonslowly varying spin textures and in systems that deviate from the adiabatic limit. Nevertheless, our observations raise a new perspective on this issue; that is, for a given spin system, whether inductor behaviour can emerge is equivalent to whether the system can store energy by applying an electric current. For instance, if the magnetic structure exhibits some elastic deformation under current, it is necessarily accompanied by some increase in the magnetic energy; accordingly, when the applied current is time varying, the energy is stored or released in response to the current variations, and this behaviour is equivalent to inductor. It could be said that there can be as many mechanisms for emergent inductors as there are mechanisms for storing energy by means of electric current. Thus, it would be an interesting direction to explore the emergent inductor function that is beyond the current-induced spin reorientation, which is the mechanism considered thus far. In general, the calculation of energy in a nonequilibrium steady state under current involves subtle issues, but elastic magnetic-structure deformations in a pinned regime appear to be well described by the Hamiltonian of an equilibrium system.

The energetic perspective discussed so far is a way of thinking that by no means allows for a negative inductance, even though the literature reports negative emergent inductance \cite{YokouchiNature, KitaoriPNAS, IedaPRB, Kurebayashi, KitaoriPRB}. For our conclusion to be coherent, we have to explain this apparent contradiction while maintaining our standpoint that physically meaningful inductance must be positive. In this context, we emphasize that in the standard equivalent circuit analysis, the observation of negative $\Im Z(\omega)$ proportional to $\omega$ does not imply negative $L$, especially when $\Re Z(\omega)$ is finite: This misunderstanding about the definition of inductance is at the root of the confusion. For instance, in the previous experiments \cite{YokouchiNature, KitaoriPNAS}, the authors observed the following complex impedance $Z(\omega)$ at a given current density:
\begin{equation}
\label{impedance}
Z(\omega)=R_{\rm DC}+i\omega\frac{\eta}{1+i\omega\tau} \:\:\:\:\:\: (\eta < 0),
\end{equation}
where the three parameters, $R_{\rm DC}$, $\eta$, and $\tau$, denote the DC resistance, a constant related with the magnitude of $\Im Z(\omega)$, and the time constant, respectively. Thus, the result, $\frac{Z(\omega) - R_{\rm DC}}{i\omega} = \frac{\eta}{1+i\omega\tau}$ with $\eta < 0$, was interpreted as the realization of negative inductance with a Debye-like frequency dependence. However, in terms of the standard equivalent circuit analysis, this $Z(\omega)$ [Eq.~(\ref{impedance})] is fully reproduced by an equivalent circuit shown in Fig.~4, which is comprised of three positive-valued elements, $R_{\rm a}, R_{\rm b}$, and $C$, that are chosen to satisfy $R_{\rm a}+R_{\rm b} = R_{\rm DC}$, $CR_{\rm b} = \tau$, and $CR_{\rm b}^2=-\eta > 0$. Thus, the observation of Eq.~(\ref{impedance})-type $Z(\omega)$ with negative $\eta$ is usually interpreted as the indication of a stray capacitance involved in the circuit, rather than a superficial negative inductance. In fact, we experimentally find that within a microfabricated sample, the system exhibits a background signal of $\frac{\Im Z(\omega)}{\omega} \approx -400$ n$\Omega$ s, which superficially corresponds to a (fictitious) negative inductance, $\approx$$-400$ nH (for details, see Supplementary Note 3 and Fig.~S2). Considering that the elimination of the background is generally not straightforward, it should be noted that $\Im Z(\omega)$ is prone to be affected by this relatively large negative-$L$-like signal.

\begin{figure}
\includegraphics{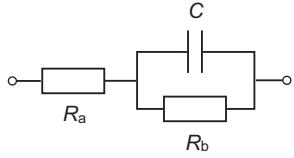}
\caption{\label{} \textbf{An equivalent circuit that can reproduce superficial negative inductance with Debye-like frequency dispersion, $\bm{Z(\omega) = R_{\rm DC}+i\omega\frac{\eta}{1+i\omega\tau}}$ with $\bm{\eta < 0}$.} The $Z(\omega)$ can be reproduced by choosing $R_{\rm a}, R_{\rm b}$, and $C$ to satisfy $R_{\rm a}+R_{\rm b} = R_{\rm DC}$, $CR_{\rm b} = \tau$, and $CR_{\rm b}^2= |\eta|$.}
\end{figure}


To conclude, we propose an energetic definition of the self-inductance coefficient, $L$, in the low-frequency regime for so-called emergent inductors and investigate its validity numerically for the case of helical magnetic textures in a pinned regime. The inductance defined from the energy increase of the magnetic system under current and that from the emergent electric field are found to agree with each other within the numerical errors, especially for the case of slowly varying spin textures and $\beta = 0$. Although our numerical approach appears to be less justified for finite $\beta$ and nonslowly varying spin textures, we conclude that the main concept of inductors in which energy is stored and released under a time-varying electric current should hold for any spin-based inductor. Conversely, if a magnetic system is capable of storing energy under current by changing the magnetic texture, the system potentially behaves as an inductor. Toward a microscopic understanding of emergent inductors, a comprehensive consideration of not only the emergent electric field but also energy will be important. Additionally, the emergent electric field beyond the linear response regime is an interesting subject, which may be more relevant to the experiments reported thus far.

\section*{D\lowercase{ata availability}}
The data used in this work are available from the corresponding author upon reasonable request.

\subsection*{A\lowercase{CKNOWLEDGMENTS}}
	The authors thank Y. Liu, S.~Maekawa, N.~Nagaosa, and G.~Tatara for their valuable discussions. S.F. and F.K. thank A. Kikkawa for providing MnSi crystals. This work was partially supported by JSPS KAKENHI (Grants No.~20K03810, No.~18H05225, No.~21H04442, and No.~23K03291), JST CREST (Grants No.~JPMJCR1874 and No.~JPMJCR20T1), and the JSPS Summer Program 2020, who funded S. H. M's placement at RIKEN, Japan.

\section*{A\lowercase{uthor contributions}}
S.F., S.H.M., and K.K. conducted the numerical calculations. S.F. analyzed the data and conducted the analytic calculation. S.F. and F.K. conducted the experiments on MnSi. F.K. conceived the project and wrote the draft with S.F. and W.K. All the authors discussed the results and commented on the manuscript.

\subsection*{C\lowercase{ompeting interests}}
The authors declare no competing interests.

\newpage

\renewcommand{\figurename}{Fig.~S}
\setcounter{figure}{0}


\section*{S\lowercase{upplementary} N\lowercase{ote} 1: I\lowercase{nstability of negative inductance}}

In previous studies on the emergent inductor, negative values of inductance, $L$, have sometimes been reported in the low-frequency limit [10--14]. As mentioned in the main text, it is known that complex impedance at low frequencies can generally be analyzed by assuming a suitable circuit consisting of positive real-valued circuit elements, $R, L$ and $C$. Here, we discuss the stability of negative $L$ by referring to the $R$--$L$ series circuit connected with a external voltage source. The circuit equation is given by:
\begin{equation}
\label{CircuitEq}
V_{\rm EX} = RI(t) + L\frac{{\rm d}I(t)}{{\rm d}t},    \tag{S1}
\end{equation}
where $V_{\rm EX}$ represents the external voltage source. For the steady-state solution, $I_{\infty} = V_{\rm EX}/R$, to be stable, the real part of solution(s) of the characteristic equation of Eq.~(\ref{CircuitEq}) should be negative; here, the characteristic equation is given by $L\lambda + R = 0$, and hence, the solution is $\lambda = -R/L$. Thus, regarding the sign of $L$, algebraic consideration concludes that a $R$--$L$ series circuit is stable when $L$ is positive, whereas it is unstable when $L$ is negative. In other words, when analyzing the low-frequency impedance $Z(\omega)$ using a $R$--$L$ series circuit, the obtained $R$ and $L$ values must both be positive real numbers. If the resulting $L$ is negative, it means that the experimental results have been analyzed using an unstable circuit, contradicting the fact that the measurements were made on a stable system. This contradiction merely indicates that the equivalent circuit used in the impedance analysis is not appropriate. As discussed in Supplementary Note 3, when the negative $L$ is obtained by using a $R$--$L$ series circuit, a $R$$\parallel$$C$ circuit should be considered as an equivalent circuit.

This conclusion does not change even if one considers, for instance, a more generalized form of a linear differential equation, such as:
\begin{equation}
\label{CircuitEq_HigherOrder}
V_{\rm EX} = RI(t) + \sum_{k=1}^\infty L_k\frac{{\rm d}^kI(t)}{{\rm d}t^k},     \tag{S2}
\end{equation}
where higher-order derivatives, $L_k\frac{d^kI}{dt^k}$, are considered for the sake of generality. For the steady-state solution to be stable, the real part of all solutions of the characteristic equation, $\sum_{k=1}^\infty L_k\lambda^k + R = 0$, should be negative: This is the conclusion from the theory of dynamical systems [36]. The necessary conditions that should be satisfied by $L_k$ and $R$ for the real part of all solutions to be negative have been mathematically answered, and they are known as the Routh-Hurwitz stability criterion [37]. This mathematical theorem tells us that one of the necessary conditions is that any $L_k$ should have the same sign as $R$; that is, all $L_k$ should be positive. This conclusion immediately indicates that the sign of $L_1$, which represents the inductance response at low frequencies, should be positive. The discussion in the main text concerns the inductance $L$ in the low-frequency regime, that is, corresponding to $L_1$ in Eq.~(\ref{CircuitEq_HigherOrder}). As we have shown, this quantity can be associated with energy storage as a result of the current application even for the case of the emergent inductor, and it should therefore be positive. Thus, the conclusion derived from the energetic point of view is consistent with the Routh-Hurwitz stability criterion.

Finally, note that the Routh-Hurwitz stability criterion does not exclude the possibility of negative $\Re L^{*}(\omega)$ at higher frequencies. In fact, substituting $I = I_0e^{i\omega t}$ into Eq.~(\ref{CircuitEq_HigherOrder}), one obtains the expression of the complex inductance $L^*(\omega)$ in terms of $L_k$ as follows:
\begin{equation}
L^*(\omega) = (L_1 -\omega^2 L_3 + \cdots) + i(\omega L_2 - \omega^3 L_4 + \cdots).    \tag{S3}
\end{equation}
This expression explicitly indicates that the real part of $L^*(\omega)$ may be negative at finite frequencies.

\section*{S\lowercase{upplementary} N\lowercase{ote} 2: A\lowercase{nalytic approach to the emergent inductance of a pristine helical magnetic texture}}

Previous literature [9, 12] has analytically derived the current-induced dynamics of a helical magnetic texture and the resulting emergent electric field (EEF) along the $x$ axis, $e_x$. In contrast, the current-induced energy variations have not been explicitly discussed, and thus, the relationship between the EEF and energy has remained unclear. In this section, after reviewing the analytic expressions derived in the literature, we discuss the energetic perspective on the emergent inductance.

Below we analytically derive the emergent inductance under an AC current along the $x$ axis for the $yz$ helical plane magnetic structure with the helical $\bm{q}$-vector parallel to the $x$ axis. The local magnetization vector of the helical magnetic order, $\bm{M}(\bm{r}) = M\bm{m}(\bm{r})$, is expressed as:
\begin{equation}
    \frac{\bm{M}(\bm{r})}{M} = (\cos\theta (\bm{r}), \sin\theta (\bm{r}) \cos\phi (\bm{r}), \sin\theta (\bm{r}) \sin\phi (\bm{r})),  \tag{S4}
\end{equation}
with $\theta (\bm{r}) = \frac{\pi}{2}$ and $\phi (\bm{r}) = qx$.

We consider the Hamiltonian, $\mathscr{H}$, consisting of the exchange interaction and the Dzyaloshinskii-Moriya (DM) interaction:
\begin{align}
    \mathscr{H} &= \int \frac{{\rm d}^3r}{a^3}\left[\frac{J^{\prime}}{2}(\nabla\bm{M})^2+D^{\prime}\bm{M}\cdot(\nabla\times\bm{M})\right] \notag \\
&=\int \frac{{\rm d}^3r}{a^3} \left[ \frac{J}{2}(\nabla\bm{m})^2 + D\bm{m}\cdot(\nabla\times\bm{m})\right],    \tag{S5}
\end{align}
where $J^{\prime}M^2=J (>0)$ is the exchange stiffness, $D^{\prime}M^2=D$ is the DM-interaction constant, and $a$ is the lattice constant. Under the electric current, $\theta$ and $\phi$ of the spiral magnetic structures vary with time. To describe the current-induced dynamics, it is helpful to introduce the collective coordinates, $\psi$ and $X$:
\begin{equation}
    \theta = \frac{\pi}{2}-\psi(t),\quad \phi = q(x-X(t)),     \tag{S6}
\end{equation} 
where $\psi$ is the tilting angle of the magnetization toward the +$x$ direction measured from the $yz$ plane and $X$ is the translational displacement of the helix along the $x$ axis. Substituting them into the Hamiltonian, one obtains:
\begin{align}
    \mathscr{H} &= \int \frac{{\rm d}^3r}{a^3}\left[\frac{J}{2}q^2\cos^2\psi-Dq\cos^2\psi \right].    \tag{S7}
\end{align}
By further considering $q = D/J$ and the sample dimension with the cross-section area $A$ (the $yz$ plane) and the length $\ell$ (along the $x$ axis), the Hamiltonian can be rewritten as:
\begin{equation}
    \mathscr{H} = -\frac{A\ell}{a^3}\frac{D^2}{2J}\cos^2\psi.    \tag{S8}
\end{equation}
The impact of an electric current on the magnetic texture arises through the $s$-$d$ coupling. The effective Lagrangian $\mathscr{L}$ and Reyleigh function $\mathscr{R}$ under electric current are represented by [12]:
\begin{align}
    \mathscr{L} &= \int \frac{{\rm d}^3r}{a^3}\hbar M(\mathcal{D}_{t}\phi)(1-\cos\theta) - \mathscr{H} - U_{\mathrm{pin}}(\bm{r})    \tag{S9}\\   
    \mathscr{R} &= \int \frac{{\rm d}^3r}{a^3} \frac{\hbar M\alpha}{2}(\mathcal{D}_{t}^{\beta}\bm{m})^2,   \tag{S10}
\end{align}
where $\mathcal{D}_{t} = \partial_{t} - \frac{Pa^3}{2|e|M}\bm{j} \cdot \nabla$, $\mathcal{D}_{t}^{\beta} = \partial_{t} - \frac{\beta}{\alpha}\frac{Pa^3}{2|e|M}\bm{j} \cdot \nabla$, and $U_{\mathrm{pin}}(\bm{r})$ represents a pinning potential ($P$ is the spin polarization and $\bm{j}$ is the electric current density).

For simplicity, we first consider the case of $\beta=0$ and no pinning potential ($U_{\mathrm{pin}} = 0$). The Euler-Lagrange equation, $\frac{\rm d}{{\rm d}t}\frac{\partial \mathscr{L}}{\partial \dot{Q}}-\frac{\partial \mathscr{L}}{\partial Q}=-\frac{\partial \mathscr{R}}{\partial \dot{Q}} \:\: (Q=\psi, X)$, for $\psi\ll1$ thus leads to 
\begin{align}
    \label{eq-1}
    &\dot{\psi} + \alpha q\dot{X} = 0,  \tag{S11}\\
    \label{eq-2}
    &q\dot{X}-\alpha\dot{\psi} = \frac{D}{\hbar M}q\psi - \frac{Pa^3}{2|e|M}qj.  \tag{S12}
\end{align} 
Here, we consider the linear-response regime under the application of an AC current $je^{i\omega t}$; that is, $|\psi|\ll\pi$ and $|\psi| \propto |j|$. Solving Eqs.~(\ref{eq-1}) and (\ref{eq-2}) leads to 
\begin{align}
   \psi(\omega) &= \zeta q j(\omega)\frac{\alpha}{\alpha qv_c + i\omega(1+\alpha^2)},  \tag{S13}\\
    X(\omega) &=\zeta j(\omega) \frac{-1}{\alpha qv_c + i\omega(1+\alpha^2)},  \tag{S14}
\end{align}
where $\zeta = \frac{Pa^3}{2|e|M}$ and $v_c = \frac{D}{\hbar M}$. Thus, one obtains the solutions of the low-frequency limit: 
\begin{equation}
    \psi(t) \approx \frac{\zeta}{v_c}j(t) = \frac{P\hbar}{2|e|}\frac{a^3}{D}j(t), \quad X(t) \approx -\frac{1}{\alpha q}\frac{P\hbar}{2|e|}\frac{a^3}{D}j(t).  \tag{S15}
\end{equation}

Using these solutions, the EEF along the $x$ axis can be calculated as follows;
\begin{align}
    e_x &= \frac{P\hbar}{2|e|} \bm{m}\cdot(\partial_x\bm{m}\times\partial_{t}\bm{m})\notag\\
        &= \frac{P\hbar}{2|e|}\sin\theta \: (\partial_x\theta \: \partial_{t}\phi-\partial_x\phi \: \partial_{t}\theta)\notag\\
        &\approx \frac{P\hbar q}{2|e|}\frac{\zeta}{v_c} \frac{{\rm d}j}{{\rm d}t}.  \tag{S16}
\end{align}
Thus, one obtains $L_{\rm {EMI}}$:
\begin{align}
    &L_{\rm {EMI}} \frac{{\rm d}I}{{\rm d}t} = e_x \ell\notag\\ 
    &\Longleftrightarrow L_{\rm {EMI}} = \frac{\ell}{A} \left( \frac{P\hbar q}{2|e|}\frac{\zeta}{v_c} \right),  \tag{S17}
\end{align}
as given in the literature [9, 12]. Similarly, $L_{\rm {energy}}$ can be calculated from the energy increase under current $I$:
\begin{align}
    &\frac{1}{2}L_{\rm {energy}}I^2 = \mathscr{H}|_{\psi=\psi(j)}- \mathscr{H}|_{\psi=0}\notag\\
    &\Longleftrightarrow L_{\rm {energy}} = \frac{2}{I^2}\times \frac{A\ell}{a^3}\frac{D^2}{2J}(1-\cos^2\psi)\notag\\
    &\Longleftrightarrow L_{\rm {energy}} = \frac{\ell}{A} \left( \frac{P\hbar q}{2|e|}\frac{\zeta}{v_c} \right).  \tag{S18}
\end{align}
Thus, for the case of a pristine helical magnetic structure with no pinning potential, $L_{\rm {EMI}} = L_{\rm {energy}}$ can be analytically shown; that is, the energy increase accompanying the current-induced magnetic-texture-distortion explains the emergent inductance. 

For quantitative estimates of the emergent inductance, another expression is more convenient. By substituting $\zeta$ and $v_c$ into the normalized inductance, or ``inductivity'', $\tilde{L} = \frac{A}{\ell}L = \frac{P\hbar q}{2|e|}\frac{\zeta}{v_c}$, one obtains:
\begin{equation}
\label{eq:L}
\tilde{L} = \left(\frac{P\hbar}{2|e|}\right)^2\frac{a^3}{J}.  \tag{S19}
\end{equation}
In our simulation, $P=1$ and $J/(2a^3)=1.8\times10^{-11}$ J m$^{-1}$. Thus,
\begin{equation}
    \tilde{L} = \left(\frac{1.054\times10^{-34}\;{\rm {J\:s}}}{2\times1.602\times10^{-19}\;{\rm {C}}}\right)^2\frac{1}{3.6\times10^{-11}\;{\rm {J\:m^{-1}}}} = 3.006\times 10^{-21}\;{\rm {H\:m}}.  \tag{S20}
\end{equation}
The typical sample dimensions used in the previous experiments are $\ell = 10^{-5}$ m and $A = 10^{-12}$ m$^2$. Hence, the expected magnitude of the linear-response emergent inductance in such a microfabricated sample is on the order of 10$^{-13}$--10$^{-14}$ H, which appears too small to detect in experiments. To observe the emergent inductance in the linear-response regime, one would have to consider the enhancement factor due to the spin-orbit interaction, which is discussed in the literature [12], or alternatively, to consider non-linear regime at large $j$, in which a large enhancement of the signal may occur [10, 11, 14].


Following the literature [12], we next derive the emergent inductance for the case of finite $\beta$ in the presence of a finite pinning potential. To use the language of the collective coordinates, we consider a specific uniform pinning potential $U_{\mathrm{pin}}$ with respect to the $X$ coordinate as follows;
\begin{equation}
    U_{\mathrm{pin}} = \int \frac{{\rm d}^3r}{a^3}w_{\mathrm{pin}}q^3X^2 = \frac{A\ell}{a^3}w_{\mathrm{pin}}q^3X^2.  \tag{S21}
\end{equation}
The Euler-Lagrange equation for $\psi\ll1$ thus leads to:
\begin{align}
    \label{eq:19}
    \dot{\psi}+\alpha q\dot{X} &= -\beta\zeta qj-v_{\mathrm{pin}}q^2X,  \tag{S22}\\
    \label{eq:20}
    \alpha\dot{\psi}-q\dot{X} &=-v_cq\psi+q\zeta j,  \tag{S23}
\end{align}
where $v_{\mathrm{pin}}=\frac{w_{\mathrm{pin}}}{\hbar S}$. For $j=j(\omega)e^{i\omega t}$, the solutions of Eqs.~(\ref{eq:19}) and (\ref{eq:20}) are:
\begin{align}
    \psi(\omega) &= \zeta q j(\omega)\frac{i\omega(\alpha-\beta)+qv_{\mathrm{pin}}}{q^2v_cv_{\mathrm{pin}} - \omega^2(1+\alpha^2)+i\omega\alpha q(v_c+v_{\mathrm{pin}})}  \tag{S24}\\
    X(\omega) &=\zeta j(\omega) \frac{-i\omega(1+\alpha\beta)-\beta qv_{c}}{q^2v_cv_{\mathrm{pin}} - \omega^2(1+\alpha^2)+i\omega\alpha q(v_c+v_{\mathrm{pin}})}.  \tag{S25}
\end{align}
In the low-frequency limit, the solutions are given by
\begin{equation}
\label{BetaSolutions}
    \psi(t) \approx \frac{\zeta}{v_c}j(t) = \frac{P\hbar}{2|e|}\frac{a^3}{D}j(t), \quad X(t) \approx -\beta\frac{P\hbar}{2|e|}\frac{a^3}{w_{\mathrm{pin}}q}j(t).  \tag{S26}
\end{equation}
According to the previous theoretical studies [32--35], the EEF should be corrected when $\beta$ is finite, as follows. 
\begin{align}
\label{BetaEEF}
    e_x &= \frac{P\hbar}{2|e|}\bm{m} \cdot (\partial_x\bm{m}\times\partial_{t}\bm{m}) - \beta\frac{P\hbar}{2|e|}(\partial_x\bm{m}\cdot\partial_{t}\bm{m}).  \tag{S27}
\end{align}
By substituting Eq.~(\ref{BetaSolutions}) into Eq.~(\ref{BetaEEF}), one obtains
\begin{align}
    e_x &= \frac{P\hbar}{2|e|}\sin\theta \: (\partial_x\theta \: \partial_{t}\phi -\partial_x\phi \: \partial_{t}\theta) - \beta\frac{P\hbar}{2|e|} (\partial_x\theta \: \partial_{t}\theta+\sin^2\theta \: \partial_x\phi \: \partial_{t}\phi)\notag\\
        &= \frac{P\hbar}{2|e|} (q\dot{\psi} + \beta q^2\dot{X}).  \tag{S28}
\end{align}
Hence, the analytic form of $L_{\mathrm{EMI}, \beta}$, which represents a value calculated according to Eq.~(\ref{BetaEEF}), is given by
\begin{equation}
\label{L_EMI_beta}
    L_{\mathrm{EMI}, \beta} = \frac{\ell}{A}  \left( \frac{P\hbar}{2|e|} \right)^2 \left( \frac{a^3}{J} -\beta^2q\frac{a^3}{w_{\mathrm{pin}}} \right),  \tag{S29}
\end{equation}
demonstrating that the correction term, $-\beta\frac{P\hbar}{2|e|}(\partial_x\bm{m} \cdot \partial_{t}\bm{m})$, reduces the EEF; i.e., $L_{\mathrm{EMI}, \beta} <  L_{\mathrm{EMI}})$. 
Contrastingly, $L_{\mathrm{energy}}$ is given as
\begin{align}
\label{L_energy_beta}
    &\frac{1}{2}L_{\mathrm{energy}}I^2 = \frac{A\ell}{a^3}\frac{D^2}{2J}\psi^2+\frac{A\ell}{a^3}w_{\mathrm{pin}}q^3\frac{X^2}{2}\notag\\
    &\Longleftrightarrow L_{\mathrm{energy}} = \frac{\ell}{A} \left( \frac{P\hbar}{2|e|} \right)^2 \left( \frac{a^3}{J}+\beta^2q\frac{a^3}{w_{\mathrm{pin}}} \right),  \tag{S30}
\end{align}
and the correction term is found to increase the energy of the magnetic system for a given current. Thus, within the present framework, the energy-conservation law does not hold for finite $\beta$ because of the presence of the correction term. Interestingly, within the present theoretical framework, the impacts of the correction term on $L_{\mathrm{EMI}}$ and $L_{\mathrm{energy}}$ are the same in magnitude, differing only in sign. However, to the best of the authors' knowledge, there is no previous theoretical study reporting $+\beta\frac{P\hbar}{2|e|}(\partial_x\bm{m} \cdot \partial_{t}\bm{m})$ in Eq.~(\ref{BetaEEF}). 

Figure~S\ref{beta} displays our simulation results for various helical magnetic textures that include random disorder, comparing (i) $\tilde{L}_{\mathrm{energy}}$ defined by the current-induced increase of the magnetic-system energy, (ii) $\tilde{L}_{\mathrm{EMI}}$ calculated from $e_x = \frac{P\hbar}{2|e|} \bm{m}\cdot(\partial_x\bm{m}\times\partial_{t}\bm{m})$, and (iii) $\tilde{L}_{\mathrm{EMI}, \beta}$ calculated from Eq.~(\ref{BetaEEF}). Although the analytic expressions are derived for the specific uniform pinning potential, the overall tendency is consistent with Eqs.~(\ref{L_EMI_beta}) and (\ref{L_energy_beta}).

\clearpage
\begin{figure}
\includegraphics{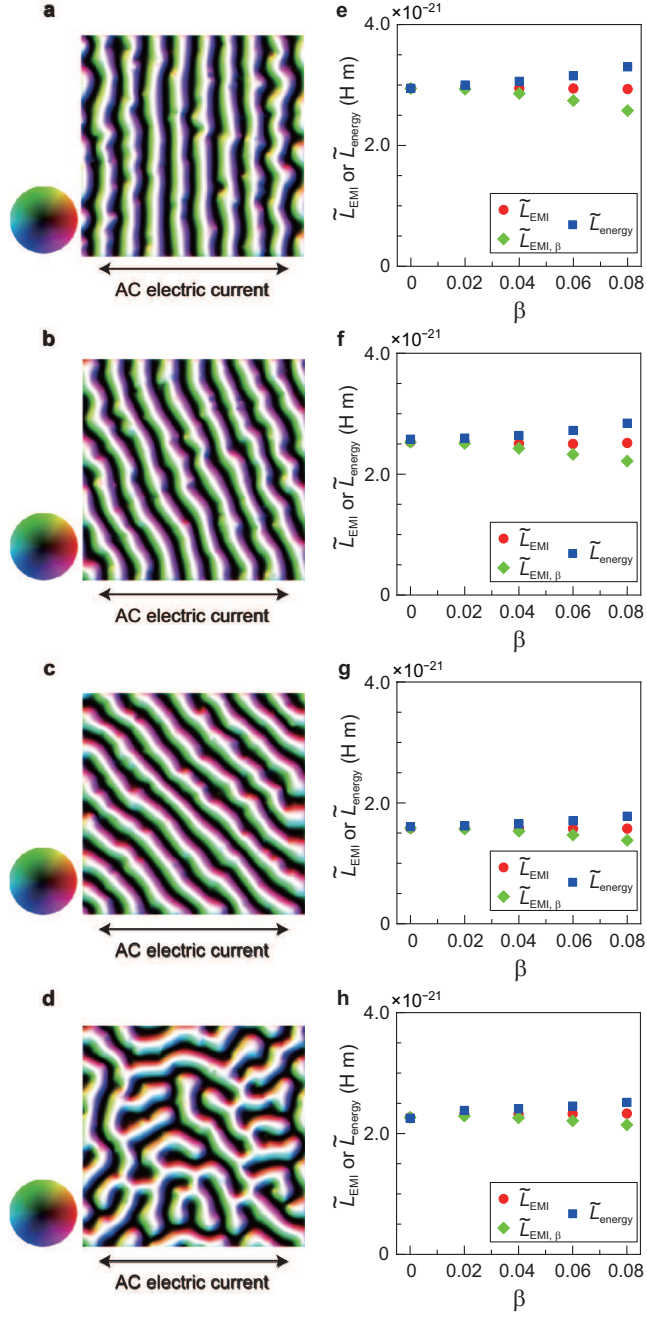}
\caption{\label{beta} \textbf{Various metastable helical textures (a--d) and corresponding inductivity (e--h).} The considered magnetic textures are the same with those discussed in the main text, and the data of $\tilde{L}_{\rm {EMI}}$ and $\tilde{L}_{\rm {energy}}$ are also the same with those shown in Fig.~3 in the main text. $\tilde{L}_{\rm {EMI}, \beta}$, which is not presented in the main text, is the value derived from the EEF given by Eq.~(\ref{BetaEEF}) .}
\end{figure}

\clearpage

\section*{S\lowercase{upplementary} N\lowercase{ote} 3: B\lowercase{ackground impedance of a microfabricated sample}}

In the main text, we have shown that if the experimentally observed $Z(\omega)= R_{\rm DC}+i\omega\frac{\eta}{1+i\omega\tau}$ with $\eta < 0$ can be reproduced by the equivalent circuit shown in Fig.~4, $Z(\omega)$ of which is given by $Z(\omega)= (R_{\rm a}+ R_{\rm b}) - i\omega\frac{R_{\rm b}^2C}{1+i\omega(R_{\rm b}C)}$. Thus, at least within the standard equivalent circuit analysis, the experimentally observed $\Im Z(\omega)$ reflects a $R$$\parallel$$C$ parallel circuit with a positive $C$, not a $R$--$L$ series circuit with a negative $L$. Note that even if this stray capacitance is a constant value, it can give rise to a temperature- and magnetic-field-dependent $\Im Z(\omega)$ through the coupling to the resistance of the sample.

To gain insight into the actual coupling between the material and the measurement system, we measured $Z(\omega)$ of the microfabricated MnSi at room temperature (Fig.~S\ref{MnSi}a). Note that the helical magnetic transition temperature of MnSi is $\approx$30 K [8], and MnSi at room temperature can therefore be regarded as an ordinary paramagnetic metal with small sample dimensions. Figure S\ref{MnSi}b,c shows the unprocessed results of $Z(\omega)$ measured with a four-terminal pair configuration using a LCR meter (as used in the previous experiments [10, 11]). We find that the $\Im Z(\omega)$ is approximately proportional to frequency with a negative slope. Although maybe misleading, the high- and low-frequency impedance can be reproduced by considering a superficial negative inductance of $-195$ nH (above 100 kHz; Fig.~S\ref{MnSi}b) and $-420$ nH (below 10 kHz; Fig.~S\ref{MnSi}c), respectively, and the characteristic crossover frequency in this representation is $\approx$30 kHz. This imaginary-part response, which is present even at room temperature, should be regarded as a background signal. Remarkably, this extrinsic background signal is larger than the experimentally derived values, such as $\approx$$-40$ nH with the characteristic frequency of $\approx$30 kHz [10]. This observation indicates that the data processing to eliminate the background signal needs a special care. Considering the analogy with a $R$$\parallel$$C$ circuit (although the actual equivalent circuit appears to be more complicated), even larger $\frac{\Im Z(\omega)}{\omega}$ may appear; for instance, if the sample resistance is 200 $\Omega$ and the stary capacitance is 500 pF, the $\frac{\Im Z(\omega)}{\omega}$ amounts to $-20$ $\mu$H in the language of fictitious negative inductance. Furthermore, given that perfectly eliminating the large background signal is challenging, the characteristic frequency involved in the background signal (in the present case, $\approx$30 kHz) may remain in the processed $\frac{\Im Z(\omega)}{\omega}$. These concerns raise a possibility that the characteristic frequency of $\frac{\Im Z(\omega)}{\omega}$ reported in the previous experiment, 10$^3$--10$^4$ Hz [10, 11], may not necessarily be of the magnetic-texture origin.

Given the fact that the data processing required to eliminate the large background signal is not straightforward, and that the experiments exclusively discuss the non-linear impedance response [10, 11, 14], we do not think that we are currently at the stage of comparing our numerical results with the previous experimental results. Rather, it would be more appropriate to discuss the theory of linear response and the experimental results independently. In fact, the linear-response theory without the enhancement factor due to the spin-orbit coupling [12] predicts the emergent inductivity of 10$^{-21}$--10$^{-19}$ H m, whereas the experimental results on microfabricated samples reports $\pm$10$^{-8}$--10$^{-5}$ H, which corresponds to $\pm$10$^{-15}$--10$^{-12}$ H m. In addition to the sign problem, a huge difference that amounts to five orders of magnitude lies between the linear-response theory and the experiments on the nonlinear regime.

\begin{figure}
\includegraphics{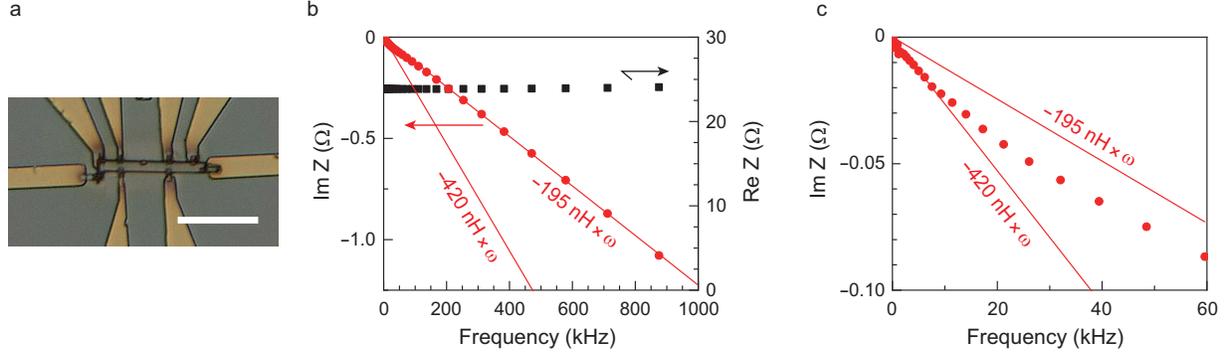}
\caption{\label{MnSi} \textbf{Impedance of a microfabricated MnSi at room temperature.} \textbf{a} Digital microscope image of the microfabricated MnSi. The scale bar is 25 $\mu$m. \textbf{b, c} Impedance of the microfabricated MnSi at room temperature: up to 1 MHz (b) and an enlarged view up to 60 kHz (c). Frequency-dependent impedance of fictitious negative inductance, $-420$ and $-195$ nH, are displayed, each of which reproduces the low- and high-frequency impedance, respectively. The crossover frequency between $-420$ and $-195$ nH is $\approx$30 kHz.}
\end{figure}
\end{document}